\documentclass[sigconf]{acmart}

\renewcommand\footnotetextcopyrightpermission[1]{} %
\acmDOI{10.475/123_4}

\acmISBN{123-4567-24-567/08/06}

\acmConference[]{McMaster}{Start}{Coding}
\acmYear{2021}
\copyrightyear{2021}

\acmArticle{4}
\acmPrice{15.00}

\usepackage{caption}
\usepackage{subcaption}

\begin{document}

\title[Code and Structure Editing for Teaching]{Code and Structure Editing for Teaching: A Case Study in using Bibliometrics to Guide Computer Science Research}

\author{Maryam Hosseinkord}
\orcid{}
\affiliation{%
  \institution{McMaster University}
  \city{Hamilton}
  \state{Ontario}
  \country{Canada}
}
\email{kordm@mcmaster.ca}

\author{Gurleen Dulai}
\orcid{}
\affiliation{%
  \institution{McMaster University}
  \city{Hamilton}
  \state{Ontario}
  \country{Canada}
}
\email{dulaig@mcmaster.ca}

\author{Narges Osmani}
\orcid{}
\affiliation{%
  \institution{McMaster University}
  \city{Hamilton}
  \state{Ontario}
  \country{Canada}
}
\email{osmanin@mcmaster.ca}

\author{Christopher K. Anand}
\orcid{}
\affiliation{%
  \institution{McMaster University}
  \city{Hamilton}
  \state{Ontario}
  \country{Canada}
}
\email{anandc@mcmaster.ca}

\renewcommand{\shortauthors}{Kord, Dulai, Osmani and Anand}

\begin{abstract}
Structure or projectional editors are a well-studied concept among researchers and some practitioners.  They have the huge advantage of preventing syntax and in some cases type errors, and aid the discovery of syntax by users unfamiliar with a language.  This begs the question: why are they not widely used in education?  To answer this question we performed a systematic review of 57 papers and performed a bibliometric analysis which extended to 381 papers.  From these we generated two hypotheses: (1) a lack of empirical evidence prevents educators from committing to this technology, and (2) existing tools have not been designed based on actual user needs as they would be if human-centered design principles were used.  Given problems we encountered with existing resources to support a systematic review, and the role of bibliometric tools in overcoming those obstacles, we also detail our methods so that they may be used as a guide for researchers or graduate students unfamiliar with bibliometrics. In particular, we report on which tools provide reliable and plentiful information in the field of computer science, and which have insufficient coverage and interoperability issues.   
\end{abstract}



\maketitle

\section{Introduction}
A graduate student or researcher approaching a new research area may use various approaches to understand the literature, including talking to colleagues, looking for a review paper, or using keyword searches in databases. But these are least likely to work for emerging and understudied areas. A systematic review provides an alternative approach. This structured approach uses pre-defined search, appraisal, synthesis, and analysis methods~\cite{grant2009typology}. Systematic research in computer science does not have universal guidelines, unlike, for example, Cochrane reviews in medicine. Arguably, the structure of information changes faster in computer science than in older branches of science, so a systematic approach needs to be better at identifying new subfields. On the other hand, computer science has provided many analytical tools to other branches of science, so it should be able to help the new graduate student who is invested in understanding the existing literature so they can make a meaningful contribution. In this paper we investigate the interconnections between research on projectional editors, other editing tools and teaching. We structure the paper to serve as a case study in how free and ad-hoc bibliometric tools can be used to accelerate the understanding of a new field.  

In this case study we seek to identify previous work on structure editors used in an educational environment and related areas of research. Structure editors allow the developer to edit the underlying structure of code (rather than relying on a compiler to validate and transform a text document) and were introduced by \citeauthor{donzeau1975structure} in 1975 in ``A structure-oriented program editor: a first step towards computer assisted programming''. Several publications followed, recounting their experience designing MENTOR, a code editing environment for a programming language called MENTOL~\cite{Donzeau-Gouge1984128}. They coined the widely used term ``structure editor''~\cite{stallman1978surveyor}. The term ``projectional editor'' gained traction following its use by Fowler in 2008. Voelter's use of this term in the paper “Embedded software development with projectional language workbenches” introduced this term to the academic literature, although ``structure editor'' is still widely used~\cite{voelter2010embedded}. Projectional editors allow the user to directly alter the abstract syntax tree, circumventing the use of a parser. Unrestricted language composition and flexible representation are inherent to projectional editing, supporting the alignment of certain programming languages with their respective domain and establishing an element of model-driven development~\cite{berger2016efficiency}. Despite the idea dating back to the 1970s and the benefits offered to novice and experienced programmers, programming predominantly relies on editing textual code, using features such as syntax highlighting and code folding~\cite{berger2016efficiency}. Empirical evidence suggests that the user experience offered by current projectional editors prevents their widespread use~\cite{voelter2014towards}. Our goal for this case study is to understand the development of research in this area. To guide our investigation, we identified four broad research questions:

\newpage
\begin{itemize}
\item[\textbf{RQ1}] How does research in computer science education, structure editors, and code editors intersect?
\item[\textbf{RQ2}] Are ideas presented in structure editor or code editor research used in computer science education?
\item[\textbf{RQ3}] Are problems faced by computer science educators being studied by researchers in these fields?
\item[\textbf{RQ4}] Given that syntax and type errors are a huge barrier to novice programmers, why are structure editors used so infrequently in computer science education, despite their initial development in 1975?
\end{itemize}

\section{Method}

\subsection{Search queries}
Identifying appropriate search queries is an important step in reviewing a research area as it will guide resource collection. Researchers should familiarize themselves with common terminology, including term variants and technical equivalents, informally before conducting a formal search. ``Wildcards'', which search word variants, can assist in this. Some search engines and databases support this type of search, but others do not. In our case study, ``projectional editor'', ``structure editor'', and ``code editor'' are the three main search terms that we used. Searching for only one of these terms would have painted an incomplete picture in the analysis phase. To avoid a similar situation, we recommend first skimming papers to become familiar with the vocabulary used in the field, making a list of definitions of commonly used terms and synonyms. The fastest way to do this is to record keywords as they are encountered while traversing the citation graph. This quickly leads to an explosion of keywords, but a bibliometric tool can help refocus the search by showing which papers have the largest number of connections to other papers in your search and particularly which papers have many connections to large clusters.  Our case study involved informally reading papers to identify important search terms, and although we discovered the use of bibliometrics to guide the search, it did help us identify foundational papers using `structure editor'' and not `projectional editor''. 

To summarize, using keyword search alone can easily discover more papers than you can reasonably read, but bibliometric-based clustering will visually show you which papers are central to your field. You can read papers from each subfield, starting with the central paper in each cluster. 

Again, in our case, structure editors differ in implementation from what a code editor is perceived to be, but their applications and purposes intersect. In response, we decided to include ``code editor'' as a search term, prompting us to re-evaluate our original first research question to include code editors. In our study, we hypothesized that papers related to code editors in general may reveal information about the usability of a structure editor. We investigated the presence of the key search terms within our selected group of papers, and noticed the use of specific terms coinciding with specific times. This pattern emphasizes the importance of identifying and using technical equivalents and closely associated terms as search queries for a given topic.

\subsection{Resource collection}
Multiple platforms were surveyed to collect relevant publications and bibliometric data, each with advantages and disadvantages, some of which were specific to our case study topic. In general, one should query multiple platforms to ensure a complete and reliable retrieval of literature. Most of these platforms support advanced search features and exporting of the search results for bibliometric management or analysis tools. 

The \textbf{Google Scholar} search engine has a large number of indexed publications, not only in scientific venues but also in online publishing platforms and patents. It supports advanced search based on wildcards, venues and search restrictions to specific website domains. Disadvantages of Google Scholar concern the absences of a native result export tool and expert cataloguing, subjecting search results to lack of precision, e.g., returning posters, blog posts and other non-peer-reviewed publications. We began our resource collection on Google Scholar before meeting with a librarian who pointed this flaw out to us. Searches in Google Scholar return an abundant amount of information, which can help develop initial familiarity with a topic, but should not be considered in a formal review. For instance, Google Scholar may be used to identify important search terms for a topic, and variations of these search terms.  

\textbf{Scopus}\footnote{\url{https://www.scopus.com/}} and \textbf{Web of Science}\footnote{\url{https://login.webofknowledge.com/}}: Initially, we used Scopus to retrieve the list of publications related to our case study and their reference data. Both Scopus and Web of Science support web-based advanced search options and many data export formats. They have indexed a large number of publications in the natural sciences, engineering, and mathematics. The exports of Scopus and Web of Science can be directly imported into the VOSviewer\footnote{\url{https://www.vosviewer.com/}} bibliometric analysis tool. When considering our case study topic, Web of Science proved to be inadequate as, compared to other tools, it had limited publications on the topic. Scopus returned a larger number of results than Web of Science using the same search queries, but upon exporting bibliographic data we discovered errors and unintended omissions in the references column. These inaccuracies became evident upon citation analysis in VOSviewer, where the generated graph was sparse. Further, data exports are limited to 2,000 works.   

We used \textbf{DBLP}\footnote{\url{https://dblp.org/}} and \textbf{ACM Digital Library} (DL)\footnote{\url{https://dl.acm.org/}} to capture a reliable list of publications. Both of these sources exclusively feature work related to computer science. ACM supports advanced search, but DBLP does not as of June 2021. Unlike DBLP, ACM does not support the export of bibliometric data, but the export of data in DBLP is limited to 1000 papers. AMiner\footnote{\url{http://aminer.org}} (sometimes referred to as Arnetminer) maintains a ``dump file'' that includes all publications indexed by DBLP, ACM, Microsoft Academic Graph, and other sources. A notable characteristic of AMiner’s publication extract is that each publication has a unique identifier, and the references field of each publication is provided in terms of these identifiers. This makes citation network generation an easy and reliable task.

\subsection{Screening}

Post resource collection, screening can be used to curate a relevant subset of research. Screening may be done paper by paper by a reader or group of readers, or by automation which involves the filtering of papers using keywords. Screening exists to avoid having irrelevant works considered in research analysis.
 
Non-automated screening poses an advantage over automation as it supports quality assurance, up to the standards of the researchers. In our case study, we used a set of 57 papers from Scopus recovered by advanced search terms “code edit*”, “projectional edit*”, “structured edit*”, “structural edit*”, and “structure edit*” and a subject filter of “Computer Science”. We screened the papers using predetermined inclusion criteria (IC), which surveyed paper abstracts, and (weighted) quality criteria (QC), concerning the paper body, to assess which papers we could include in analysis.

\begin{itemize}
\item[\textbf{IC1}] The study’s main focus is code editors.
\item[\textbf{IC2}] The study presents empirical evidence or implementation details.
 
\item[\textbf{QC1}] The study clearly states the aim of research. (1 point)
\item[\textbf{QC2}] The study is contextualized in terms of other studies and research. (1 point)
\item[\textbf{QC3}] The study justifies system, algorithmic, and design decisions. (1 point)
\item[\textbf{QC4}] The study describes methods used in detail, such that the experiment could be reproduced. (1 point)
\item[\textbf{QC5}] The performance metrics of the study are rationalized. (0.5 points)
\item[\textbf{QC6}] Statistical analysis is done on the findings. (0.5 points)
\item[\textbf{QC7}] Findings are supported by empirical evidence. (0.5 points)
\end{itemize}

Some of these criteria existed on a scale. For instance, when considering QC6, we ranked those with no empirical evidence as a 0, those with empirical evidence and insufficient statistical analysis as a 0.5, and those with empirical evidence and strong statistical analysis as a 1. Twenty-nine papers satisfied the inclusion criteria and had a quality rating of at least 4.5 out of 5.5, the predetermined threshold decided on by screeners, and were included in the  analysis. All papers were independently evaluated by two screeners, and the seven discrepancies in score were discussed and resolved.
 
Automation bypasses the need for screeners to read every single paper to be included in the analysis. As a result, automation can be faster and may be more suitable for large data sets. To compensate for the absence of a search tool for DBLP, we implemented a Python script to screen the ``dump file'' of all publications indexed by DBLP, maintained and constantly updated by Arnetminer~\cite{tang2008arnetminer}. The latest ``dump file'' as of June 2021 was extracted on May 14, 2021, and contains more than 5M papers and 48M links. The Python script reads and filters papers in the dataset based on our original search queries. A total of 381 relevant papers were recovered from the DBLP data set using automation. Only papers containing the original search queries in the title, abstract, and keywords were kept for analysis, and the rest were discarded. Depending on the research topic or goal, one may decide to automate based on the title, abstract, or keywords, or a combination of them all. If the main focus of a paper is education, then it is likely that this keyword would be found in the title, abstract, or keywords of the paper. In this case, if a false negative was lost to automation, the paper would likely be unfocused and should be screened out for quality. For bibliometric analysis, false negatives could result in the absence of a significant graph node. Contrarily, false positives are not as detrimental to bibliometric analysis, as clustering will discard them. One may choose to automate more conservatively or liberally depending on their research topic, route of analysis, and whether false positives or false negatives are of higher concern.

In the case of automated screening, inclusion criteria and quality criteria involve splitting, or ``black-and-white'' characterization. Automation of QC6 above is not likely to be a simple automation to implement without an existing framework. Although possible, automation of this quality criteria may warrant more time spent than desired on the screening component of a review. A researcher may find that a combination of non-automated and automated screening may be the most efficient way of screening. In the case of our research topic, screening by reading was required for some criteria, such as level of empirical evidence. Although criteria such as IC1 could be done by hand, using automation is likely more suitable, especially when considering large data sets.

\subsection{Analyzing Data (Systematic Review)}
Following screening, analysis methods were used to understand the current state of the research topic. Patterns that arise may indicate gaps in a topic or suggest further avenues of research.
 
In traditional reviews, screeners read papers to understand their purpose in the context of the research area. Using the collection of screened papers from Scopus, we used tagging to contextualize the research. We categorized the papers using one or more tags:

\begin{itemize} 
\item[\textbf{Tag 1}] The paper provides design principles for code editors.
\item[\textbf{Tag 2}] The paper talks about the software quality of code editors.
\item[\textbf{Tag 3}] The paper talks about a necessary feature of code editors.
\item[\textbf{Tag 4}] The paper evaluates editors in the context of education.  
\item[\textbf{Tag 5}] The paper presents a new structure or code editor.
\item[\textbf{Tag 6}] The paper presents a structure editor generator.
\end{itemize}
 
The number of papers tagged may suggest answers to our research questions. Of the 29 selected papers, three papers were tagged using Tag 1 and one paper was tagged using Tag 4. Through tagging, it became evident that the reason why structure editors are seldom used in education may be because not enough research has been published in the area. This could be the result of poorly defined design principles for structure editors, which are essential to user-centered design. Publications regarding code editor frameworks and features outnumber those concerning user experience. This suggests that structure editor features may be potentially valuable, but their delivery to the user is preventing their use.
 
Aside from tagging, screeners may pick up on patterns in the research as they read all the papers in the screening phase. In our case study, one example involved the use of text colour/brightness in code editors. \citeauthor{beckmann2020efficient} presented a new tree-oriented structure editor, acknowledging that structure editors pose usability issues to users as structure editors involve a different editing metaphor than text editors [1]. One way that this editor accommodates this new metaphor is by using brighter text colours to indicate deeper nesting in the code. The use of brighter colours to indicate nesting seems arbitrary as this choice is not supported by empirical data. A different paper develops and tests an algorithm to predict meta-features of a syntax theme based on the colour data of the theme~\cite{sterling2018patterns}. RGB colour data yields accuracy greater than 0.73 in predicting the theme’s metadata, including the language it’s intended for~\cite{sterling2018patterns}. Although syntax is not of concern in structure editors, no papers regarding the development of a new structure editor considered text colour as a way of increasing usability by familiarity, specifically for experienced programmers who want to switch from a text editor to a structure editor. Although tagging and reading the papers may point to a common area, having insufficient and disjointed research in design principles and usability of code editors, it can be difficult to understand how significant these gaps are in the topics being considered. This gap could be attributed to the fact that one paper focused on a structure editor and the other on text editors. By reading alone, it can be hard to know if these research areas are disjointed, or if they connect in ways that do not involve design principles.

\subsection{Analyzing Data (Bibliometrics)}
A student may choose to use automated tagging based on keywords to categorize their papers. Reading a substantial set of papers without automation can be a difficult and time-consuming task, especially when considering a large number of papers, such as the set we recovered using AMiner. 

In our case study, we were looking to understand why structure editors, despite their potential to help programmers overcome barriers, have not been accepted or commercialized. Through analysing the smaller group of papers from Scopus, we considered that a lack of design principles for structure editors could be preventing the widespread adoption of structure editors. We tagged our large set of papers from AMiner by searching for words or phrases within the title, abstract, and keywords of each paper. Tags such as ``agile'' and ``Design Thinking'' were easy to use as these words do not have many variants or equivalents. Finding a tag for teaching or addressing novice developers requires more thought, given that these words have many variants and equivalents. To find papers related to the aforementioned tags, we also used “edu*”, “teach*”, and “tutor*” as search terms. When analysing the number of papers with each tag, we noticed that few papers focus on user-centric methods to design structure editors, which could explain the poor usability and, consequently, user experience. 

Bibliometric tools can be used to support the understanding of the literature of an overarching research area. Within our case study, we explored the use of VOSviewer, Gephi, and Python programs to analyse our bibliometric exports. These tools support analysis types including co-authorship, co-occurrence, citation, bibliographic coupling, and co-citation within a group of papers. Of these, building a citation network is a common way of understanding how a collection of papers reference each other, which may indicate how important some papers are within the specific field of study. Each vertex in a citation network is a paper, and a directed link from paper $A$ to paper $B$ shows a citation from $A$ to $B$. Hubs with a high number of edges are associated with papers that are referenced the most. 

\begin{figure}[h]
  \begin{subfigure}[b]{\linewidth}
    \includegraphics[width=\linewidth]{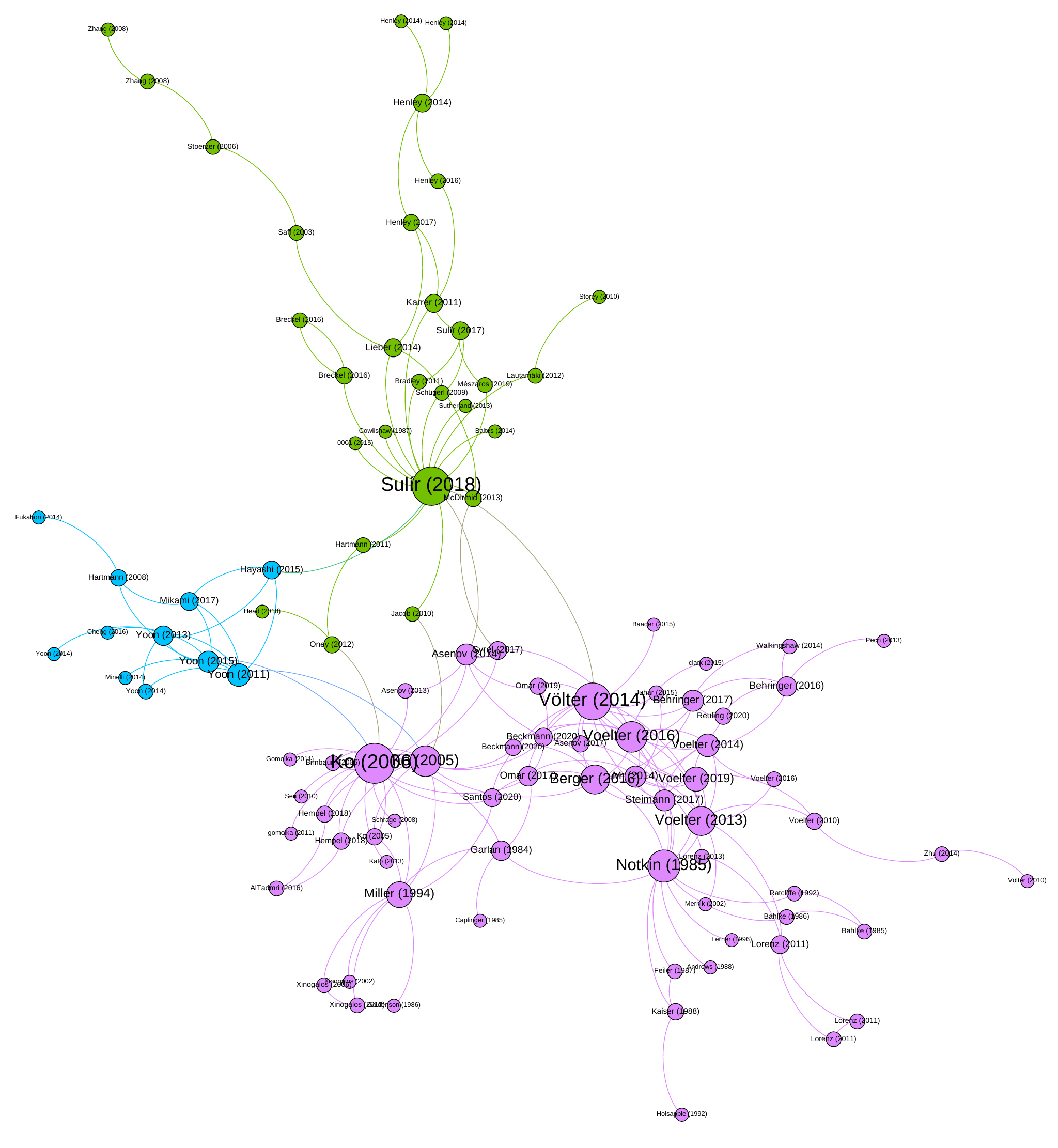}
    \label{fig:citation-network-sub}
  \end{subfigure}
  \vfill
  \begin{subfigure}[b]{\linewidth}
    \includegraphics[width=\linewidth]{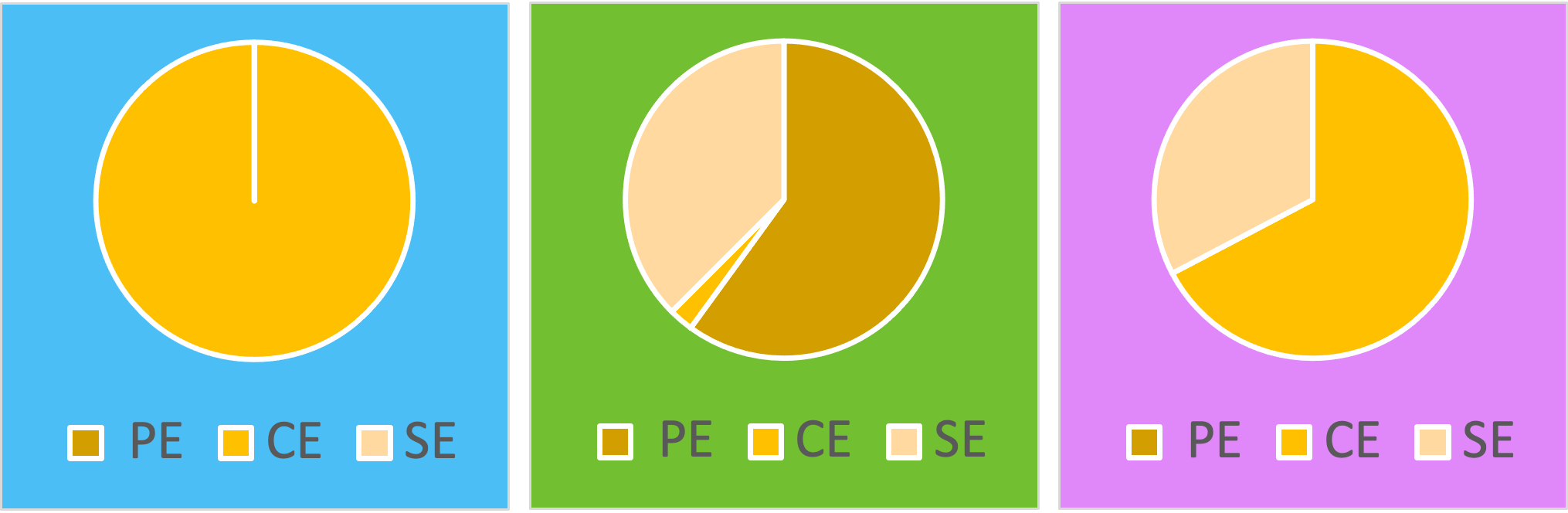}
    \label{fig:cls2-pie}
  \end{subfigure}
  \caption{\textit{Top:} Citation network of papers related to structure editors, projectional editors, and code editors. The vertices are the papers and the links show the citations among papers. The links are curved so that the direction is indicated by clockwise traversal. \textit{Bottom:}  The number of publications of each class in each cluster.}
  \label{fig:citation-network}
\end{figure}

Figure~\ref{fig:citation-network} shows the network we have obtained using the screened collection of papers from AMiner. Numerous analytical studies can be conducted on the citation network, including the cluster or community detection analysis. The community detection algorithms try to find highly interconnected areas in the network. In the case of a citation network, these areas would be a collection of papers that cite each other. So, papers within a cluster are related. The connection pattern between the papers in each cluster is locally significant compared to papers outside that cluster.

We used Gephi to process citation network graphs. For clustering, we used the Louvain algorithm with \textit{modularity}~\cite{Blondel_2008} as the quality function. Modularity is a measure of how much the connectivity pattern of a graph (or subgraph) differs from a random graph in which that pattern is expected to be uniform. Thus, it can be used to specify which partitioning of the graph is better than others. Louvain clustering uses the strength of vertex interconnections to cluster the network. The resolution parameter~\cite{Lambiotte_2014} in Gephi’s implementation is initially set to $1.0$. A higher resolution enables the algorithm to find smaller clusters of highly connected nodes, and hence, results in a higher number of clusters. Setting a smaller resolution will find a smaller set of final clusters. Every researcher can tweak this parameter to obtain the desired number of clusters based on the research area. Since we aimed to obtain three clusters, potentially relevant to three types of editors we searched for, we tried setting different values for \textit{resolution} using $vertex$ to obtain the final three clusters in purple, green and blue shown in Figure~\ref{fig:citation-network}. To test this hypothesis, we compared the clusters to keywords appearing in papers in the three clusters. The results are shown as pie charts in Figure~\ref{fig:citation-network}. 
We find that there is a small exclusively code-editor cluster of 11 papers, but the larger clusters are mixed.  The next cluster of 39 papers contains both projectional and structure editor papers.
This makes sense, because projectional and structure editors are essentially the same idea.
The largest cluster of 48 papers, however, does not refer to projectional editors, only the older terms “structure editor” as well as “code editor”, and in fact 60\% of its papers refer to the latter.

\smallskip
This seems like a weak conclusion, but in fact it changed our perception of the field.
In the initial ACM DL and Google Scholar search and systematic review, most papers came from the middle cluster, and were dominated by papers related to the MPS system.  
Reading this part of the literature gives the impression that MPS largely invented the concept of structure editor and that the structure editor literature was disjoint from the code editor literature. 
This illustrates a problem with social networks: that popular nodes tend to become more popular, making it harder to discover other interesting nodes.  Exploring the nodes in the large cluster counters this view, with structure and code editor papers referencing common literature.  So an initial hypothesis that there is a gap in the literature due to the failure of the structure editor literature to learn from the code editor literature turns out not to be true.  If that gap were real, it might also help explain another gap, the lack of empirical evidence for proposed structure editors.  Whereas structure and code editors differ radically in their construction, empirical evaluation of their usability should share methods and objectives.  Perhaps this does somewhat explain the structure of the literature, but it is certainly not the explanation it seemed initially.

\begin{figure}[h]
  \centering
  \includegraphics[width=\linewidth]{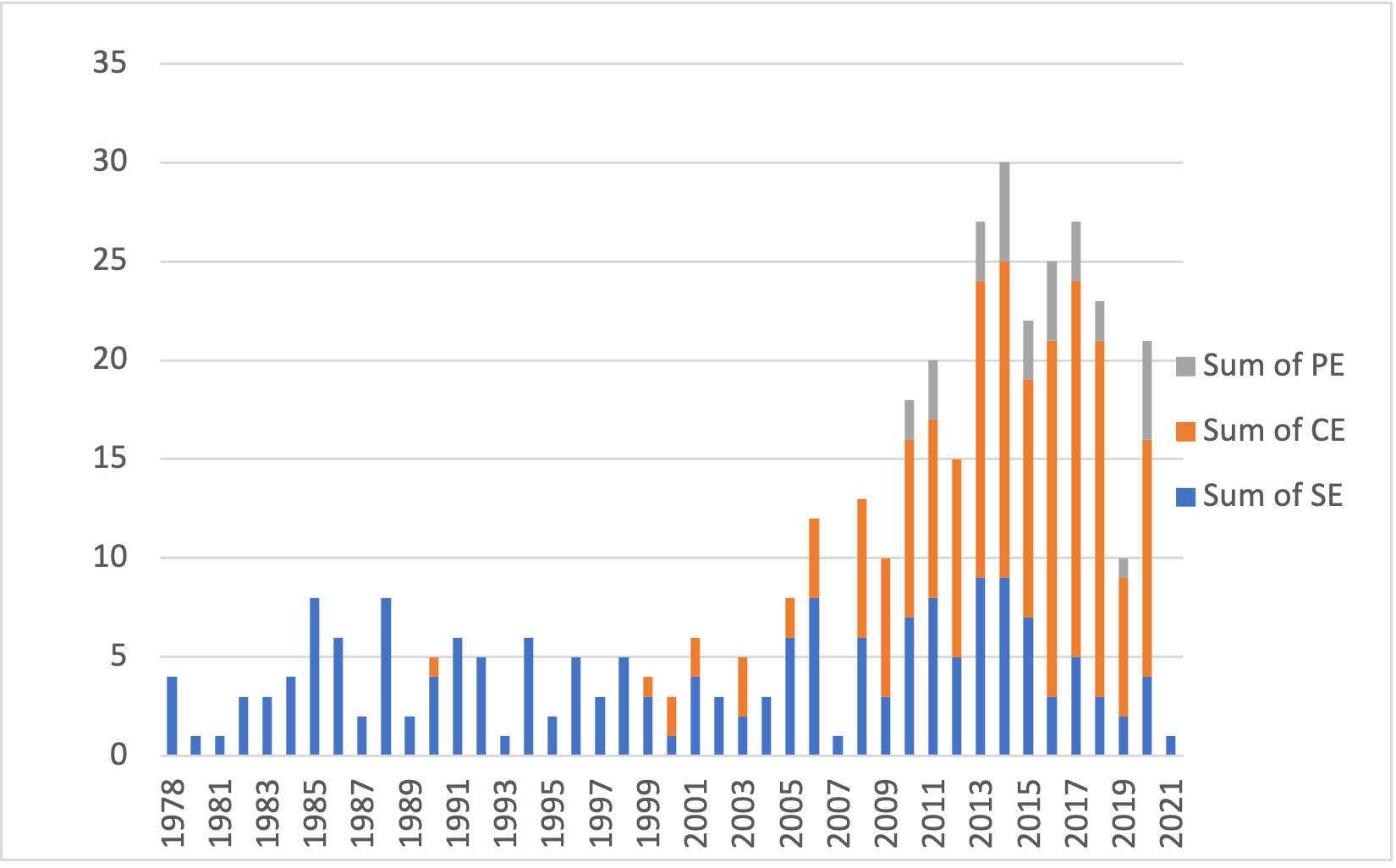}
  \caption{The number of publications over years about Projectional and Code editors.}
  \Description{A woman and a girl in white dresses sit in an open car.}
  \label{fig:pubs-over-years}
\end{figure}

Figure \ref{fig:pubs-over-years} shows the number of papers in each category for each year.

Building the citation network is a common way of understanding how a collection of papers refer to each other and what are the most important papers in a specific field of study.
Each vertex in a citation network is a paper and a directed link from paper $A$ to paper $B$ shows a citation from $A$ to $B$. The hubs, i.e., vertices with high number of edges in this network are the most important papers that many papers referenced to them.

Figure~\ref{fig:citation-network} shows the network we have obtained using the screened collection of papers from AMiner. Numerous analytical studies can be conducted on the citation network, including the cluster or community detection analysis. The community detection algorithms try to find highly interconnected areas in the network. In the case of a citation network, these areas would be a collection of papers that cite each other. So, papers within a cluster are related. The connection pattern between the papers in each cluster is locally significant compared to papers outside that cluster.

\section{Discussion and Conclusions}
\textbf{RQ1} \textit{How does research in computer science education, projectional editors, and code editors intersect?}   Our initial systematic review suggested that there was little communication between structure editor and code editor researchers. Subsequent bibliometric analysis revealed that we were overweighting a cluster of papers, whose authors used the terminology ``projectional editor'', which did not connect to the code editor literature, but there was a larger cluster who preferred the term ``structure editor'', showing that with some exceptions, most researchers in these subfields were aware of work in the other field.
 
\noindent
\textbf{RQ2} \textit{Are ideas presented in structure editor or code editor research used in computer science education?} Yes, researchers are both inspired by the ideas from the research literature, and targeting specific properties of structure editors, e.g., syntax discoverability.  But of the 381 papers, only 8 were tagged as related to education.

\noindent\textbf{RQ3} \textit{Are problems faced by computer science educators being studied by researchers in these fields?}  We found no evidence for this.  Of the papers read, most start with a solution and not an investigation of the problem, even for papers which do evaluation.  (See also the discussion of Human-Centered Design below.)

\noindent\textbf{RQ4} \textit{Given that syntax and type errors are a huge barrier to novice programmers, why are structure editors used so infrequently in computer science education, despite their initial development in 1975?} We formulated two hypotheses for this, which should be studied further, but could already inform planning of future research in this field.

 \textbf{Empirical Evidence}  We hypothesized that structure editors were not widely used in education because none of the papers in the systematic review presented empirical evidence of the impact on new users, only 3 of the 8 papers tagged as related to education %
 presented empirical evidence, and \citeauthor{xinogalos2013using}'s review of structure editors did not find any evaluated by independent researchers.  So empirical evidence is lacking, but mostly because there are so few papers related to education.
 
\textbf{Human-Centered Design} We similarly hypothesized that structure editors were not widely used in education because no papers in the systematic review described the use of human-centered design, e.g., Design Thinking, to develop \textit{usable} interfaces.  Automatic tagging also found no such papers among the 381, but rereading the GNOME paper \cite{garlan1985gnome} based on its position in the cluster graph, and with the knowledge developed during the systematic review, we think that GNOME (a structure editor for undergraduates programming Pascal and Karel) was developed using human-centered design principles even though the language of human-centered design did not yet exist, and they attributed their success to this fact.  They also made it clear that it was necessary to develop language-specific interfaces for their languages through observation and feedback from their users.  This probably accounts for the failure of their project to influence current research, as Pascal was replaced by other languages.

\subsubsection{Tool Recommendations}
Systematic reviews are labour intensive, and are usually performed by teams.  Newer reference managers have collaboration features like shared notes and highlighting which facilitate teamwork, as well as features like full-text search. We used Mendeley as a reference manager and a place to store papers pre- and post-screening. It supports the export of libraries in multiple formats, but not the analysis of this data. Scopus and Web of Science allow for readers to make a list of papers as they search through the database, but are harder to use than Mendeley and also lack analysis tools. Web of Science has poor coverage of this field, and data exports from Scopus were unreliable. As a result, we recommend AMiner as the preferred database for computer science. Although AMiner contains abstracts, and is therefore sufficient for initial screening, and has links to full papers, it does not support teamwork the way Mendeley does. To perform bibliometric analyses, we used VOSviewer, Gephi, and Python, and can also recommend these tools. From Python it is easy to search results in CSV or Excel format and use a spreadsheet program (including web-based collaboration software) for sorting and searching using keywords, recording scores and filtering by combinations of keywords, scores, publication date and cluster membership.  We used Excel integrated with Microsoft Teams.

Unfortunately, we came across no tool that integrated a database search, source management, reliable data exports, bibliometric analysis and collaboration. This is a gap which open-source software developers, publishers or other tool vendors should fill.

In other domains, evaluating empirical evidence plays a central role in systematic reviews. We know that our automatic tagging is not reliable, and do not believe it is possible to be so. One solution would be to make tagging of empirical evidence part of the publication process, with referees scoring papers on their use of empirical evidence and making source data available for review. We recommend this as a way of encouraging the use of empirical evidence and highlighting the need for such in computer science research involving human-computer interaction or otherwise impacting society.

\begin{acks}
We acknowledge financial support from IBM Centre for Advanced Studies and NSERC.
\end{acks}

\bibliographystyle{ACM-Reference-Format}
\bibliography{sample-base}


\begin{thebibliography}{14}


\ifx \showCODEN    \undefined \def \showCODEN     #1{\unskip}     \fi
\ifx \showDOI      \undefined \def \showDOI       #1{#1}\fi
\ifx \showISBNx    \undefined \def \showISBNx     #1{\unskip}     \fi
\ifx \showISBNxiii \undefined \def \showISBNxiii  #1{\unskip}     \fi
\ifx \showISSN     \undefined \def \showISSN      #1{\unskip}     \fi
\ifx \showLCCN     \undefined \def \showLCCN      #1{\unskip}     \fi
\ifx \shownote     \undefined \def \shownote      #1{#1}          \fi
\ifx \showarticletitle \undefined \def \showarticletitle #1{#1}   \fi
\ifx \showURL      \undefined \def \showURL       {\relax}        \fi
\providecommand\bibfield[2]{#2}
\providecommand\bibinfo[2]{#2}
\providecommand\natexlab[1]{#1}
\providecommand\showeprint[2][]{arXiv:#2}

\bibitem[\protect\citeauthoryear{Beckmann}{Beckmann}{2020}]%
        {beckmann2020efficient}
\bibfield{author}{\bibinfo{person}{Tom Beckmann}.}
  \bibinfo{year}{2020}\natexlab{}.
\newblock \showarticletitle{Efficient editing in a tree-oriented projectional
  editor}. In \bibinfo{booktitle}{\emph{Conference Companion of the 4th
  International Conference on Art, Science, and Engineering of Programming}}.
  \bibinfo{pages}{215--216}.
\newblock


\bibitem[\protect\citeauthoryear{Berger, V{\"o}lter, Jensen, Dangprasert, and
  Siegmund}{Berger et~al\mbox{.}}{2016}]%
        {berger2016efficiency}
\bibfield{author}{\bibinfo{person}{Thorsten Berger}, \bibinfo{person}{Markus
  V{\"o}lter}, \bibinfo{person}{Hans~Peter Jensen}, \bibinfo{person}{Taweesap
  Dangprasert}, {and} \bibinfo{person}{Janet Siegmund}.}
  \bibinfo{year}{2016}\natexlab{}.
\newblock \showarticletitle{Efficiency of projectional editing: A controlled
  experiment}. In \bibinfo{booktitle}{\emph{Proceedings of the 2016 24th ACM
  SIGSOFT International Symposium on Foundations of Software Engineering}}.
  \bibinfo{pages}{763--774}.
\newblock


\bibitem[\protect\citeauthoryear{Blondel, Guillaume, Lambiotte, and
  Lefebvre}{Blondel et~al\mbox{.}}{2008}]%
        {Blondel_2008}
\bibfield{author}{\bibinfo{person}{Vincent~D Blondel},
  \bibinfo{person}{Jean-Loup Guillaume}, \bibinfo{person}{Renaud Lambiotte},
  {and} \bibinfo{person}{Etienne Lefebvre}.} \bibinfo{year}{2008}\natexlab{}.
\newblock \showarticletitle{Fast unfolding of communities in large networks}.
\newblock \bibinfo{journal}{\emph{Journal of Statistical Mechanics: Theory and
  Experiment}} \bibinfo{volume}{2008}, \bibinfo{number}{10}
  (\bibinfo{date}{oct} \bibinfo{year}{2008}), \bibinfo{pages}{P10008}.
\newblock
\urldef\tempurl%
\url{https://doi.org/10.1088/1742-5468/2008/10/p10008}
\showDOI{\tempurl}


\bibitem[\protect\citeauthoryear{Donzeau-Gouge, Huet, Kahn, and
  Lang}{Donzeau-Gouge et~al\mbox{.}}{1984}]%
        {Donzeau-Gouge1984128}
\bibfield{author}{\bibinfo{person}{Veronique Donzeau-Gouge},
  \bibinfo{person}{Gerard Huet}, \bibinfo{person}{Gilles Kahn}, {and}
  \bibinfo{person}{Bernard Lang}.} \bibinfo{year}{1984}\natexlab{}.
\newblock \bibinfo{booktitle}{\emph{Programming Environments Based on
  Structured Editors: The MENTOR Experience.}}
\newblock 128--140 pages.
\newblock
\urldef\tempurl%
\url{https://www.scopus.com/inward/record.uri?eid=2-s2.0-0021570876&partnerID=40&md5=41fdab3d07ab5d53a61e4c4e340d390a}
\showURL{%
\tempurl}
\newblock
\shownote{cited By 62.}


\bibitem[\protect\citeauthoryear{Donzeau-Gouge, Huet, Kahn, Lang, and
  Levy}{Donzeau-Gouge et~al\mbox{.}}{1975}]%
        {donzeau1975structure}
\bibfield{author}{\bibinfo{person}{Veronique Donzeau-Gouge}, \bibinfo{person}{G
  Huet}, \bibinfo{person}{Gilles Kahn}, \bibinfo{person}{Bernard Lang}, {and}
  \bibinfo{person}{JJ Levy}.} \bibinfo{year}{1975}\natexlab{}.
\newblock \bibinfo{booktitle}{\emph{A structure-oriented program editor: a
  first step towards computer assisted programming}}.
\newblock \bibinfo{publisher}{IRIA. Laboratoire de Recherche en Informatique et
  Automatique}.
\newblock


\bibitem[\protect\citeauthoryear{Garlan and Miller}{Garlan and Miller}{1984}]%
        {garlan1985gnome}
\bibfield{author}{\bibinfo{person}{David~B. Garlan} {and}
  \bibinfo{person}{Philip~L. Miller}.} \bibinfo{year}{1984}\natexlab{}.
\newblock \showarticletitle{GNOME: An Introductory Programming Environment
  Based on a Family of Structure Editors}.
\newblock \bibinfo{journal}{\emph{SIGPLAN Not.}} \bibinfo{volume}{19},
  \bibinfo{number}{5} (\bibinfo{date}{April} \bibinfo{year}{1984}),
  \bibinfo{pages}{65–72}.
\newblock
\showISSN{0362-1340}
\urldef\tempurl%
\url{https://doi.org/10.1145/390011.808250}
\showDOI{\tempurl}


\bibitem[\protect\citeauthoryear{Grant and Booth}{Grant and Booth}{2009}]%
        {grant2009typology}
\bibfield{author}{\bibinfo{person}{Maria~J Grant} {and} \bibinfo{person}{Andrew
  Booth}.} \bibinfo{year}{2009}\natexlab{}.
\newblock \showarticletitle{A typology of reviews: an analysis of 14 review
  types and associated methodologies}.
\newblock \bibinfo{journal}{\emph{Health information \& libraries journal}}
  \bibinfo{volume}{26}, \bibinfo{number}{2} (\bibinfo{year}{2009}),
  \bibinfo{pages}{91--108}.
\newblock


\bibitem[\protect\citeauthoryear{Lambiotte, Delvenne, and Barahona}{Lambiotte
  et~al\mbox{.}}{2014}]%
        {Lambiotte_2014}
\bibfield{author}{\bibinfo{person}{Renaud Lambiotte},
  \bibinfo{person}{Jean-Charles Delvenne}, {and} \bibinfo{person}{Mauricio
  Barahona}.} \bibinfo{year}{2014}\natexlab{}.
\newblock \showarticletitle{Random Walks, Markov Processes and the Multiscale
  Modular Organization of Complex Networks}.
\newblock \bibinfo{journal}{\emph{IEEE Transactions on Network Science and
  Engineering}} \bibinfo{volume}{1}, \bibinfo{number}{2} (\bibinfo{date}{Jul}
  \bibinfo{year}{2014}), \bibinfo{pages}{76–90}.
\newblock
\showISSN{2327-4697}
\urldef\tempurl%
\url{https://doi.org/10.1109/tnse.2015.2391998}
\showDOI{\tempurl}


\bibitem[\protect\citeauthoryear{Stallman}{Stallman}{1978}]%
        {stallman1978surveyor}
\bibfield{author}{\bibinfo{person}{Richard~M Stallman}.}
  \bibinfo{year}{1978}\natexlab{}.
\newblock \showarticletitle{Surveyor's Forum: Structured Editing with a Lisp}.
\newblock \bibinfo{journal}{\emph{ACM Computing Surveys (CSUR)}}
  \bibinfo{volume}{10}, \bibinfo{number}{4} (\bibinfo{year}{1978}),
  \bibinfo{pages}{505--507}.
\newblock


\bibitem[\protect\citeauthoryear{Sterling and Suleimenov}{Sterling and
  Suleimenov}{2018}]%
        {sterling2018patterns}
\bibfield{author}{\bibinfo{person}{Mark Sterling} {and}
  \bibinfo{person}{Aidarbek Suleimenov}.} \bibinfo{year}{2018}\natexlab{}.
\newblock \showarticletitle{Patterns of syntax theme customization for code
  editors}. In \bibinfo{booktitle}{\emph{2018 IEEE 3rd International Conference
  on Communication and Information Systems (ICCIS)}}. IEEE,
  \bibinfo{pages}{173--176}.
\newblock


\bibitem[\protect\citeauthoryear{Tang, Zhang, Yao, Li, Zhang, and Su}{Tang
  et~al\mbox{.}}{2008}]%
        {tang2008arnetminer}
\bibfield{author}{\bibinfo{person}{Jie Tang}, \bibinfo{person}{Jing Zhang},
  \bibinfo{person}{Limin Yao}, \bibinfo{person}{Juanzi Li}, \bibinfo{person}{Li
  Zhang}, {and} \bibinfo{person}{Zhong Su}.} \bibinfo{year}{2008}\natexlab{}.
\newblock \showarticletitle{Arnetminer: extraction and mining of academic
  social networks}. In \bibinfo{booktitle}{\emph{Proceedings of the 14th ACM
  SIGKDD international conference on Knowledge discovery and data mining}}.
  \bibinfo{pages}{990--998}.
\newblock


\bibitem[\protect\citeauthoryear{Voelter}{Voelter}{2010}]%
        {voelter2010embedded}
\bibfield{author}{\bibinfo{person}{Markus Voelter}.}
  \bibinfo{year}{2010}\natexlab{}.
\newblock \showarticletitle{Embedded software development with projectional
  language workbenches}. In \bibinfo{booktitle}{\emph{International Conference
  on Model Driven Engineering Languages and Systems}}. Springer,
  \bibinfo{pages}{32--46}.
\newblock


\bibitem[\protect\citeauthoryear{Voelter, Siegmund, Berger, and Kolb}{Voelter
  et~al\mbox{.}}{2014}]%
        {voelter2014towards}
\bibfield{author}{\bibinfo{person}{Markus Voelter}, \bibinfo{person}{Janet
  Siegmund}, \bibinfo{person}{Thorsten Berger}, {and} \bibinfo{person}{Bernd
  Kolb}.} \bibinfo{year}{2014}\natexlab{}.
\newblock \showarticletitle{Towards user-friendly projectional editors}. In
  \bibinfo{booktitle}{\emph{International Conference on Software Language
  Engineering}}. Springer, \bibinfo{pages}{41--61}.
\newblock


\bibitem[\protect\citeauthoryear{Xinogalos}{Xinogalos}{2013}]%
        {xinogalos2013using}
\bibfield{author}{\bibinfo{person}{Stelios Xinogalos}.}
  \bibinfo{year}{2013}\natexlab{}.
\newblock \showarticletitle{Using flowchart-based programming environments for
  simplifying programming and software engineering processes}. In
  \bibinfo{booktitle}{\emph{2013 IEEE Global Engineering Education Conference
  (EDUCON)}}. IEEE, \bibinfo{pages}{1313--1322}.
\newblock


\end{thebibliography}

\end{document}